# Valley-Coupled-Spintronic Non-Volatile Memories with Compute-In-Memory Support

Sandeep Thirumala[1*], *Student Member IEEE,* Yi-Tse Hung[1,2], Shubham Jain[1], Arnab Raha[3], *Member IEEE,* Niharika Thakuria[1], Vijay Raghunathan[1], *Senior Member IEEE,* Anand Raghunathan[1], *Fellow IEEE,* Zhihong Chen[1,2], *Senior Member IEEE,* and Sumeet Gupta[1], *Member IEEE*

[1]School of Electrical and Computer Engineering, Purdue University, West Lafayette, IN
[2]Birck Nanotechnology Center, Purdue University, West Lafayette, IN; [3]Intel Corporation, Santa Clara, CA
*sthirum@purdue.edu

***Abstract*—In this work, we propose valley-coupled spin-hall memories (VSH-MRAMs) based on monolayer $WSe_2$. The key features of the proposed memories are (a) the ability to switch magnets with perpendicular magnetic anisotropy (PMA) via VSH effect and (b) an integrated gate that can modulate the charge/spin current ($I_C/I_S$) flow. The former attribute results in high energy efficiency (compared to the Giant-Spin Hall (GSH) effect-based devices with in-plane magnetic anisotropy (IMA) magnets). The latter feature leads to a compact access-transistor-less memory array design. We experimentally measure the gate controllability of the current as well as the non-local resistance associated with VSH effect. Based on the measured data, we develop a simulation framework (using physical equations) to propose and analyze single-ended and differential VSH effect based magnetic memories (VSH-MRAM and DVSH-MRAM, respectively). At the array level, the proposed VSH/DVSH-MRAMs achieve 50%/ 11% lower write time, 59%/ 67% lower write energy and 35%/ 41% lower read energy at iso-sense margin, compared to single-ended/differential (GSH/DGSH)-MRAMs. System level evaluation in the context of general purpose processor and intermittently-powered system shows up to 3.14X and 1.98X better energy efficiency for the proposed (D)VSH-MRAMs over (D)GSH-MRAMs respectively. Further, the differential sensing of the proposed DVSH-MRAM leads to natural and simultaneous in-memory computation of bit-wise AND and NOR logic functions. Using this feature, we design a computation-in-memory (CiM) architecture that performs Boolean logic and addition (ADD) with a single array access. System analysis performed by integrating our DVSH-MRAM: CiM in the Nios II processor across various application benchmarks shows up to 2.66X total energy savings, compared to DGSH-MRAM: CiM.***

## I. INTRODUCTION

Over the last decade, there has been an immense interest in emerging non-volatile memories (NVMs) due to their distinct advantages over the traditional silicon-based memories [1], such as near-zero stand-by leakage and high integration densities [2-3]. However, they possess design conflicts and issues associated with reliability, robustness and high write energy [4-8]. Spin-based memories using magnetic tunnel junctions (MTJs) [9] look promising with good endurance and high integration densities. Specifically, spin-transfer-torque magnetic RAM (STT-MRAM) has attracted immense interest. Samsung's STT-MRAM in 28nm FDSOI platform [10] and Intel's FinFET based MRAM technology [11] are some industrial efforts on the implementation of spintronic memory. However, there are several challenges which still need to be addressed. For example, they exhibit low distinguishability between their bi-stable states making them prone to sensing failures [12]. Also, due to their two-terminal cell-design, the write and read paths are coupled, leading to design challenges.

Recent advancements with the possibility of generating spin polarized current using charge current in heavy metals has led to the realization of the Giant Spin Hall (GSH) effect based MRAM [13-14] (also known as spin-orbit-torque MRAM; SOT-MRAM). Compared to STT-MRAMs, GSH-MRAM showcase significant improvement in write energy along with the possibility to independently co-optimize the read and write operations due to their decoupled read and write current paths. GSH effect also enables the possibility of achieving a differential storage due to the simultaneous generation of opposite polarized spin currents [15]. However, both the single ended and differential memory designs based on GSH effect require multiple access transistors leading to a significant area penalty [14, 15]. Also, the spin injection efficiency which is directly proportional to the spin hall angle ($\theta_{SH}$ <0.3) is low for these heavy metals [13, 16]. This results in performance degradation and energy inefficiency. Another drawback with GSH-MRAMs is that they can only switch IMA magnets without the presence of any external magnetic field or geometrical changes to the ferromagnet [17, 18]. As PMA magnets are known to be more energy efficient in switching and thermally stable than IMA [19], GSH-MRAMs offer limited performance and energy benefits. Therefore, there arises a need to explore new memory technologies to harness the full potential of spin-based storage.

On the application front, data-intensive workloads have come to the forefront in recent years. This has led to frequent and humongous number of data accesses from the memory system to the processor. As a result, larger amount of storage is required, which demands the exploration for high density memory solution. On the other hand, due to the larger delay associated with memory access compared to the processing time (also known as memory-wall problem [20]), the data movement to and from the memory cell (across the bit-lines, memory interface and interconnects) is a major performance and energy bottleneck in standard computing architectures. Therefore, there is also a need to explore alternate computing paradigms such as Computation-in-Memory (CiM), where computations are performed inside the memory array [21-22]. This reduces the data transfer between memory and processor, thereby improving the performance and energy efficiency.

Most of the prior efforts on CiM designs using spin-based information storage involve the use of single ended STT-MRAM or SOT/GSH-MRAMs [23-26]. Multi-word-line assertion along with a modified sense amplifier and peripheral circuitry enables Boolean and arithmetic operations to be performed within the memory array [23]. Although STT-CiM [23] benefits form the high density and good endurance of STT-MRAM, and GSH-MRAM based CiM [24] overcomes the drawbacks of high write energy consumption in STT-MRAMs, they both suffer from degraded robustness during in-memory compute. This is attributed to the poor distinguishability between their bi-stable states yielding deteriorated sense margins during compute operations [23, 24]. Therefore, it is important to explore robust and energy-efficient CiM designs utilizing the benefits offered by spin-based information storage for current and future generation of compute systems involving large amount of data.

To that effect, in this paper, we propose non-volatile memory devices based on Valley-coupled-Spin Hall (VSH) effect with the ability to naturally switch PMA magnets, leading to higher energy efficiency compared to GSH-MRAM. Moreover, exploiting the spin generation through a semiconductor ($WSe_2$) rather than a metal (in GSH), we propose an integrated gating in our NVM devices, which enable access transistor less bit-cell design leading to large integration density. Furthermore, the proposed devices inherently lead to differential read functionality. Leveraging this attribute, we present an array design with energy efficient computation-in-memory capabilities, which can potentially alleviate the von-Neumann bottleneck and overcome the drawbacks of existing spin-based CiM designs. The key contributions of this paper are as follows:

- We propose energy-efficient VSH effect based single-ended and differential spintronic memory devices and their access-transistor-less arrays. The single-ended and differential design are referred to as VSH- and DVSH-MRAMs, respectively. We develop a simulation framework for evaluating the proposed memories and calibrate it with experiments.
- We propose computation of Boolean logic and arithmetic addition operations within the memory array (CiM) with simultaneous assertion of multiple word-line with the proposed DVSH-MRAM.
- We design a reconfigurable current sense amplifier which can dynamically switch its operation between differential mode for memory read and single-ended sensing mode for in-memory compute in the proposed DVSH-MRAM.
- We perform detailed array and system-level analysis for the proposed (D)VSH-MRAMs in comparison with existing (D)GSH-MRAMs in the context of a general purpose processor and an intermittently-powered system.

## II. Background

### A. Giant Spin Hall (GSH) Effect

The Giant Spin Hall effect is an efficient mechanism for generating spin polarized currents. A charge current passing through a heavy metal layer such as Ta, Pt or W have been experimentally demonstrated to generate in-plane spin polarized currents [13, 16] (Fig. 1(a)). The GSH effect is mainly used for switching magnetization of IMA magnets. Deterministic switching of PMA magnets using GSH effect requires externally assisted magnetic field to break the symmetry [17] or complex design modifications to the MTJ geometry [18]. The major advantage with GSH effect-based magnetization switching is the low write current/energy when compared to the STT-based magnetization switching [13-16].

The generated spin current ($I_S$) to charge current ($I_C$) ratio which is also known as the spin injection efficiency is directly proportional to the spin hall angle, $\theta_{SH}$ [14]. Experiments have shown $\theta_{SH}$ ~ 0.1-0.3 for heavy metals, resulting in low spin injection efficiency [13, 16]. Furthermore, the efficiency of GSH effect is impacted by the spin-flip length ($\lambda_S$), which characterizes the mean distance between spin-flipping collisions. $\lambda_S$ has been calculated to be around ~1-2nm [13, 16] for heavy metals with large GSH effect.

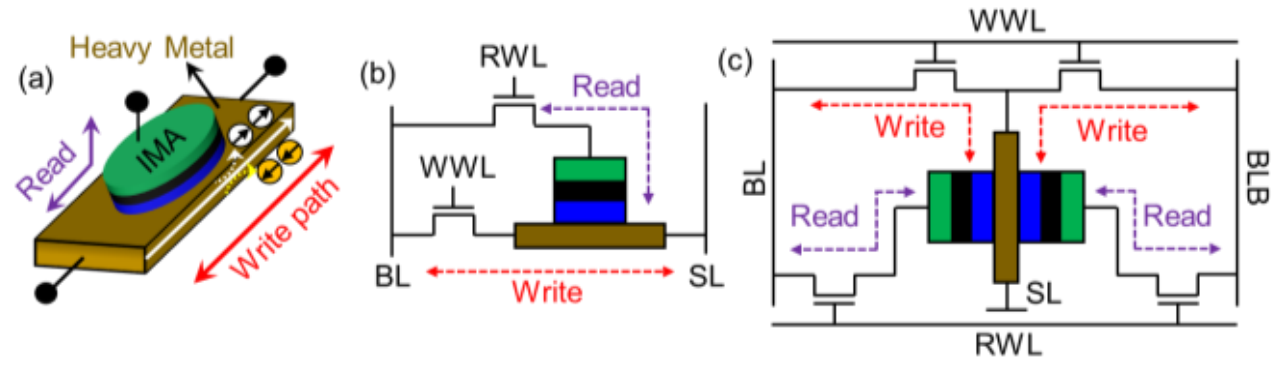

Fig. 1. (a) GSH effect in heavy metal leading to magnetization switching in MTJ. (b) Single ended GSH-MRAM and (c) Differential DGSH-MRAM bit-cell schematics.

### B. GSH Effect based non-volatile memories

The three terminal device structure of the GSH effect-based spin device (Fig. 1(a)) mitigates the read-write conflict of the two terminal STT-MRAM due to the separation of read-write paths. Moreover, such an approach has shown to be promising for energy-efficient storage compared to STT-MRAM [13-16]. Several bit-cell designs have been proposed using the GSH effect [14-15]. Fig. 1(b) shows circuit schematic of GSH-MRAM which consist of a read and write access transistor for single-ended memory [14]. Write operation is achieved by turning ON the write access transistor and depending on the direction of charge current flow (which determines the spin current polarization), the MTJ state is stored. The spin current interacting with the MTJ to deterministically switch the magnetization is calculated as:

$$I_S = \frac{A_{MTJ}}{A_{HM}} * \theta_{SH} * I_C \qquad (1)$$

where $A_{MTJ}$ and $A_{HM}$ are the cross-sectional area of MTJ and heavy metal, respectively [14]. The read operation is carried out by turning ON the read access transistor and sensing the resistance state of the MTJ (parallel (P) or anti-parallel (AP)). As the read and write paths are decoupled, they can be optimized independently [14, 27].

Utilizing the opposite spin generation at the top and bottom surfaces of the heavy metal, a differential GSH-MRAM (DGSH-MRAM) was proposed in [15] with two MTJs placed on either side of the heavy metal layer (Fig. 1(c)). This leads to true and complimentary bit storage in the memory cell. The write operation remains the same as GSH-MRAM while the read is achieved using differential sensing, leading to higher sense margins. However, compared to GSH-MRAM, two more additional transistors are required to selectively access a bit cell in an array without disturbing the unassessed cells. Furthermore, fabrication of true and differential MTJs on the top and bottom sides of the heavy metal may increase processing complexities and costs.

The above mentioned GSH effect-based memory designs have been proposed to switch IMA based MTJs. This is because, only in-plane spin polarized currents are generated in the heavy metals. IMA magnets are not suitable for ultra-scaled dimensions due the limit on the aspect ratio of the free layer as well as low thermal stability [17-19]. In comparison, PMA magnets are more stable and robust at scaled dimensions with high packing density, which is mainly attributed to the absence of in-plane shape magnetic anisotropy [27]. Moreover, due to the absence of de-magnetization fields, lower energy is required for magnetization switching in PMA magnets compared to IMA, even at iso-thermal stability [27]. Although, GSH effect based PMA switching has been demonstrated with external magnetic field [17], or a local di-polar field [28] or introducing tilted anisotropy in the ferromagnet [18], the feasibility of achieving such a design change in scaled, high density technologies is yet to be explored. Moreover, the requirement of additional access transistors for GSH effect based bit-cell designs leads to large area overheads which also increases the energy consumption for bit-line and word-line charging.

To address the aforementioned challenges associated with GSH-effect based devices, we propose to utilize the valley-coupled-Spin Hall (VSH) effect in monolayer $WSe_2$ to design

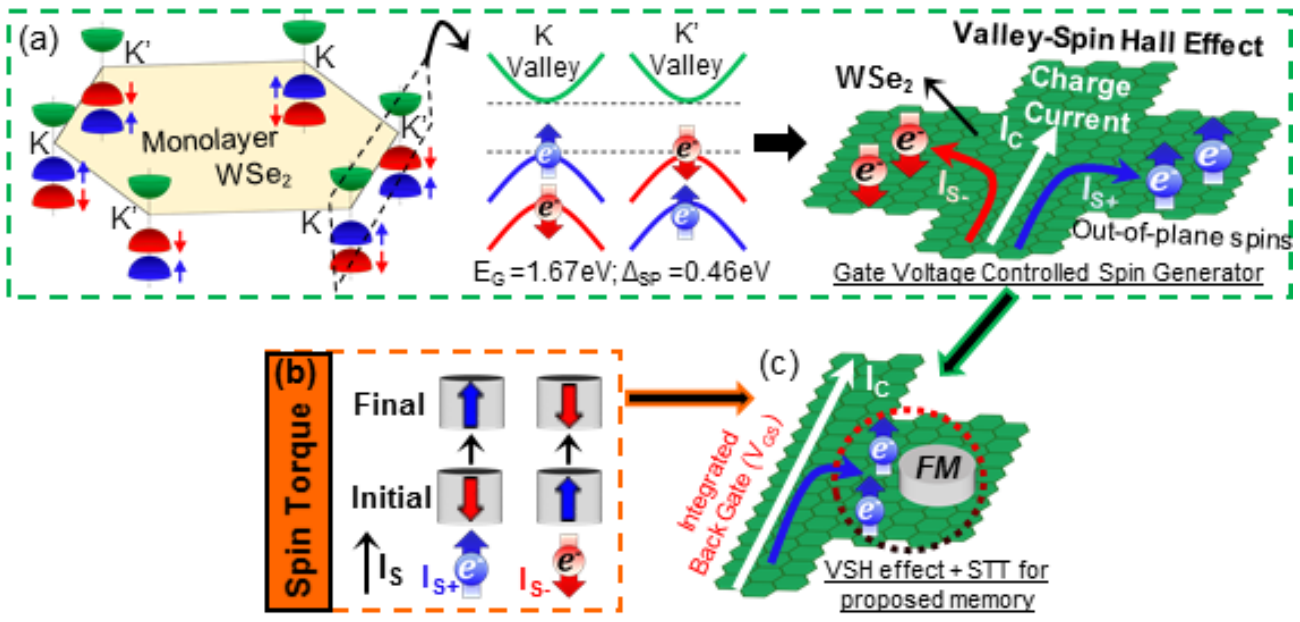

Fig. 2 (a) Band structure of $WSe_2$ showcasing spin-valley coupling resulting in VSH Effect (b) STT for switching PMA magnet. (c) Proposed idea of coupling VSH effect and spin torque for NVM design.

MRAMs based on PMA magnets. The VSH effect is naturally suited to switch PMA magnets, which promises higher energy efficiency in the proposed designs. Before we discuss our memory designs and the associated benefits and trade-offs, let us briefly review the VSH effect, next.

## *C. Valley-coupled-Spin Hall (VSH) Effect*

Monolayer transition metal dichalcogenides (TMDs) are multi-valley 2D semiconductors (Fig. 2(a)) with inherent broken inversion and preserved time reversal symmetries. As a result, carriers in the K and K' valleys of the valence band (p-type) possess nonzero Berry curvature (Ω) such that Ω(K)= -Ω(K'). The resultant transverse carrier velocity in TMDs such as $WSe_2$ with large valence band spin-splitting ($\Delta_{SP}$) [29], leads to valley-coupled spin currents on the application of electric field. This phenomenon is called the VSH effect [30-32]. VSH effect in $WSe_2$ generates out-of-plane spin polarized currents ($I_{S+}/I_{S-}$ ; Fig. 2(a)) which can switch PMA magnets without any external magnetic field ($B_{EXT}$) or complex changes to the MTJ structure, unlike GSH-effect based memory devices [17-18, 28].

It has been experimentally demonstrated that monolayer TMDs exhibit a large valley-hall angle, $\theta_{VH}$ ~1 [30] at 25°C. Now, due to the existence of strong spin-valley coupling in monolayer $WSe_2$ [31-32] (as a result of large $\Delta_{SP}$), the $\theta_{SH}$ is expected to be equal to $\theta_{VH}$, i.e., $\theta_{SH}$ ~1. The large $\theta_{SH}$ corresponds to high spin injection efficiency which can potentially lead to enhanced energy efficiency during magnetization switching. In contrast, GSH effect exhibit relatively much smaller $\theta_{SH}$ ~ 0.1-0.3. Moreover, VSH effect resulting in out-of-plane spin generation exhibits $\lambda_S$ of 0.5-1μm [30-31] (unlike GSH effect in heavy metals; $\lambda_S$ ~ 1-2nm). The large $\theta_{SH}$ and $\lambda_S$ in monolayer $WSe_2$ opens up new opportunities for information storage.

Utilizing the unique attributes of VSH effect in conjunction with spin torque physics (Fig. 2(b, c)), we propose valley-coupled spintronic devices in this paper and show their utility for energy-efficient data storage and computation in-memory. Section III involves the description of experimental demonstration of VSH effect in $WSe_2$. Simulation framework has been detailed in Section IV. Section V and VI discusses the non-volatile memory design and computation in-memory respectively, using the proposed VSH effect based memories. Section VII concludes this paper.

# III. Fabrication and Experimental Results

## *A. Device Fabrication*

The process flow for the fabrication of a Hall bar device structure (Fig. 3(a)) is illustrated in Fig. 3(b). Chemical vapor deposition (CVD) grown $WSe_2$ films were transferred to 90nm $SiO_2$ substrates with highly doped silicon on the back

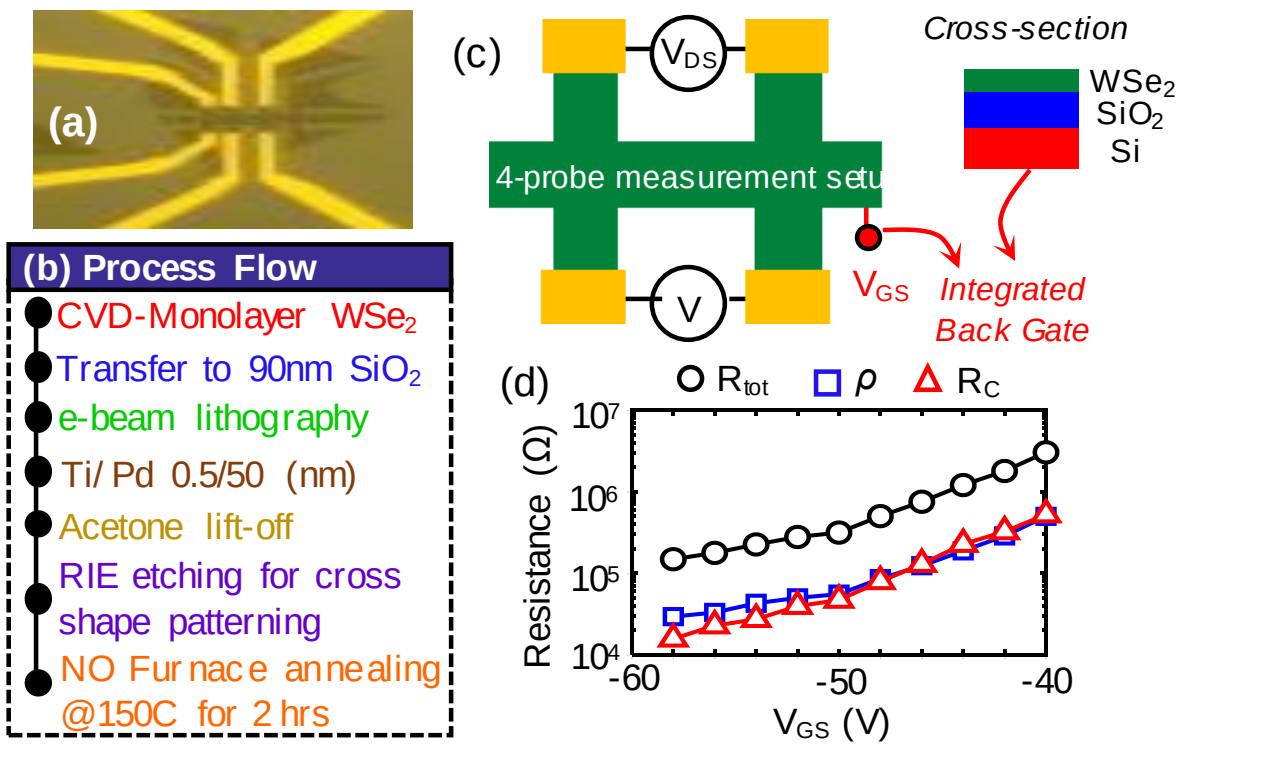

Fig. 3 (a) Optical microscope image of the hall bar device structure (b) Fabrication process flow (c) Cartoon of 4-probe measurement setup (d)Total, sheet and contact resistances versus gate voltage, $V_{GS}$.

side. Doped Siserves as the integrated back gate for controlling the flow of $I_C$ and $I_S$ (as explained later). Standard e-beam lithography using PMMA A4 950 resist was employed to pattern electric contacts on the CVD $WSe_2$ flakes. Ti/Pd (0.5/50nm) was deposited in an e-beam evaporator followed by a lift-off process in acetone. CVD grown BN film was transferred from Cu foil onto the devices through a process that involves etching the Cu foil with iron chloride ($FeCl_3$) and immersing it in diluted HCl and DI water alternatingly for few times before scooping up. This BN layer was inserted to minimize device degradation from PMMA residues after the RIE etching process. RIE etching mask was defined by e-beam lithography using PMMA A4 950 resist and BN/$WSe_2$ flakes were etched using Ar/SF6 for 10 seconds. The final devices underwent nitric oxide (NO) furnace annealing at 150°C for two hours followed by vacuum annealing (~ $10^{-8}$ torr) at 250°C for four hours to minimize PMMA residue and threshold voltage shift due to trap charges.

## *B. Parameter Extraction*

Two types of measurements were performed, as illustrated in Fig. 3 and 4. A conventional four probe measurement (Fig. 3 (c)) was conducted to extract sheet resistance ($\rho$), contact resistance ($R_C$) and total resistance ($R_{TOT}$) (Fig. 3(d)). The non-local (NL) measurements were performed to probe the Hall voltage induced by any carrier distributions due to the VSH and its reciprocal effect (Fig. 4(a)). Fig. 4(b, c) shows gate control of charge current ($I_C$) and NL resistance ($R_{NL}= V_{NL}/I_C$) for different device samples with arm lengths ($L_A$) equal to 2μm, 3μm and 5μm. $R_{NL}$ vs $L_A$ at $V_{GS}$= -60V was used to extract the spin flip length, $\lambda_S$ = 550nm from the fitting of $R_{NL} \propto e\,(-L_A/\lambda_S)$ [30-31] (Fig. 4(d)). It is important to note that the spin-generator is a p-type device and therefore it requires negative gate-to-source voltages to turn it ON.

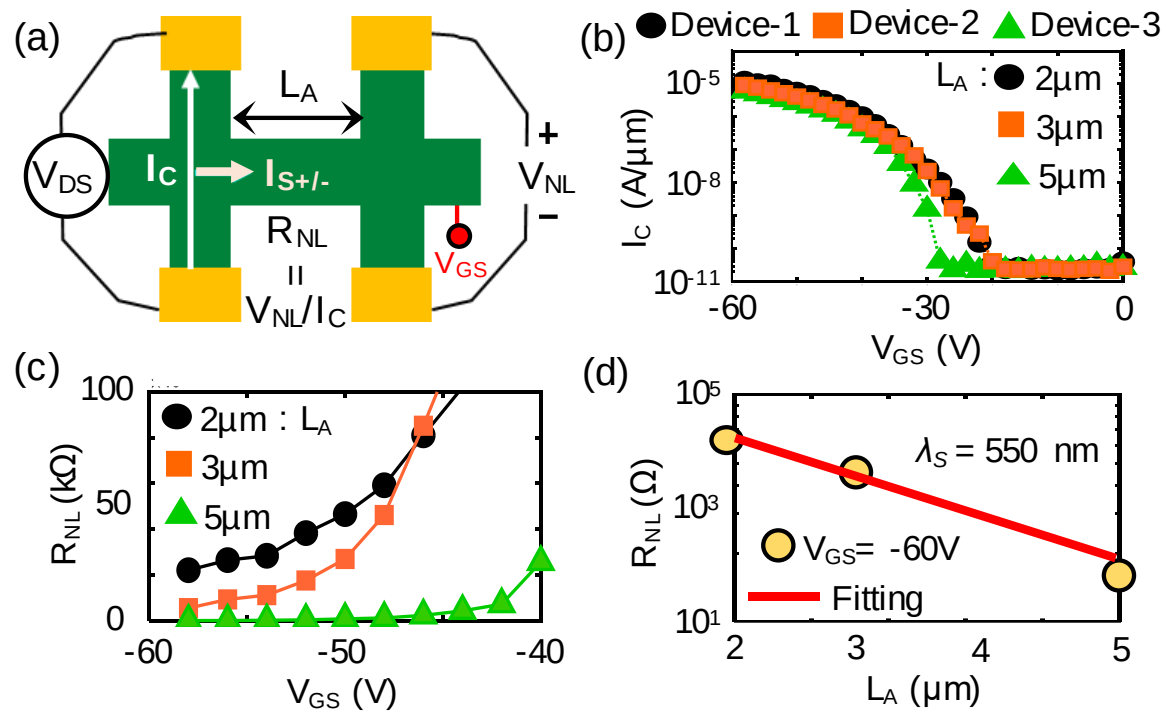

Fig. 4 (a) Non-local measurement setup (b) Charge current ($I_C$) vs $V_{GS}$ for different arm lengths, $L_A$ (c) Non-local resistance ($R_{NL}$) vs $V_{GS}$ for different $L_A$ (d) $R_{NL}$ vs $L_S$ to extract spin flip length, $\lambda_S$.

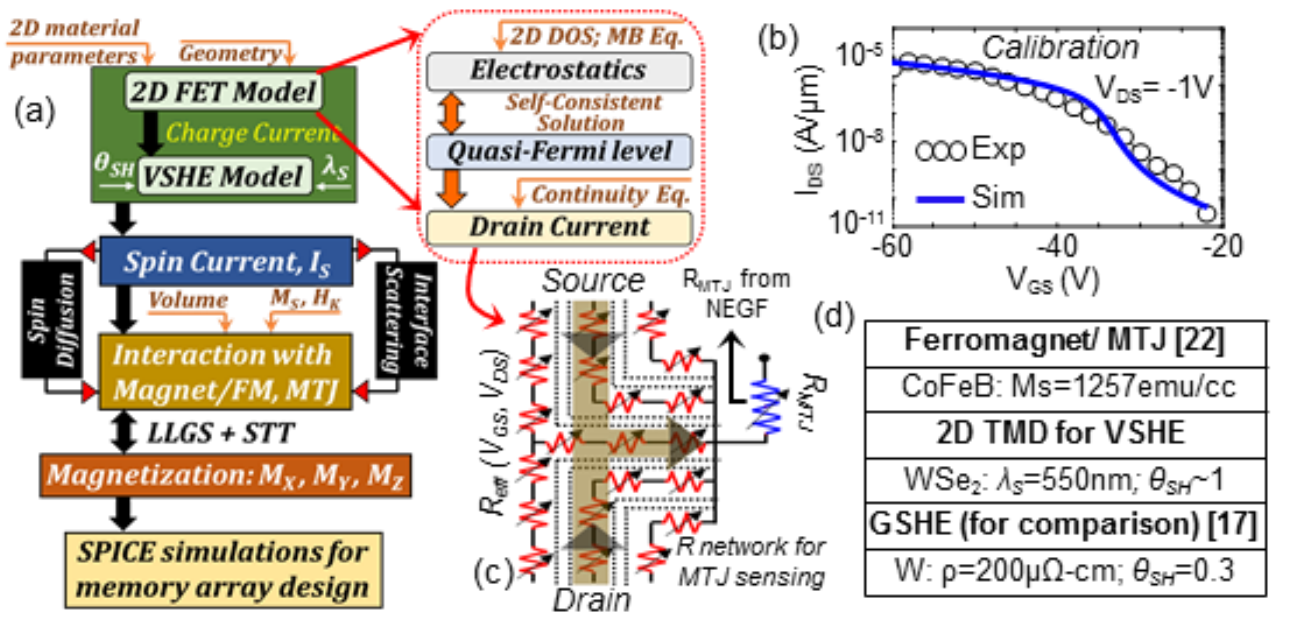

Fig. 5 (a) Self-consistent simulation framework (b) Calibration of the monolayer $WSe_2$ FET (c) SPICE-based distributed resistance network for sensing MTJ resistance (d) Material parameters

## IV. Simulation Framework

We have built a self-consistent simulation framework in SPICE for the proposed valley-coupled spintronic memory device/array evaluation (Fig. 5(a)). Monolayer $WSe_2$ electrostatics is modelled using the capacitance network model suggested in [29], albeit with modification for back-gated device used in this work. Further, we model the charge current using the continuity equations for drift-diffusion transport as proposed in [29] (calibration in Fig. 5(b)). The charge current is then used in conjunction with the Valley Spin Hall effect model, which calculates the spin current based on the experimental $\theta_{SH}$ and $\lambda_S$ values [30-31]. $I_S$ interacting with the free layer of MTJ is calculated as:

$$I_S = \frac{D_{MTJ}}{L_G} * \theta_{SH} * I_C \tag{2}$$

where $D_{MTJ}$ is the diameter of MTJ (circular) and $L_G$ is gate length of the transistor (see Fig. 6). Spin diffusion and interface scattering are considered in the monolayer $WSe_2$ channel while calculating the spin current flow as per the method proposed in [30, 34]. Landau-Lifshitz-Gilbert-Slonczewski (LLGS) equation is used to model the switching dynamics of the PMA magnet, which serves as the free layer (FL) of a magnetic tunnel junction (MTJ) formed on top of the TMD (as described later). For sensing, the MTJ resistance ($R_{MTJ}$) is obtained from NEGF [27]. Further, as we will discuss in detail later, the read path is 'T'/ 'H' shaped. To properly account for the sensed currents, we use a distributed resistive network (Fig. 5(c)) based on the conductance of $WSe_2$ layer and the shape of the read path. Therefore, the read path includes the resistance of the MTJ as well as that of conducting $WSe_2$ layer. The sensed currents are used to read the bit-information stored and also perform computation in memory, as discussed extensively in Section V and VI. We incorporate contact resistances at the drain terminal, source terminal and MTJ-TMD interface based on [35] (accounting for Schottky barrier). The contact resistances play a crucial role and require further investigation for performance/energy optimization. The simulations parameters used in this work are shown in Fig. 5(d). Note, in all of our analysis, we evaluate the proposed memory devices and circuits considering a minimum gate length ($L_G$) of 45nm (for system compatibility – see Section V.D). With the understanding of the simulation methodology, let us now present the proposed VSH effect based spintronic memory.

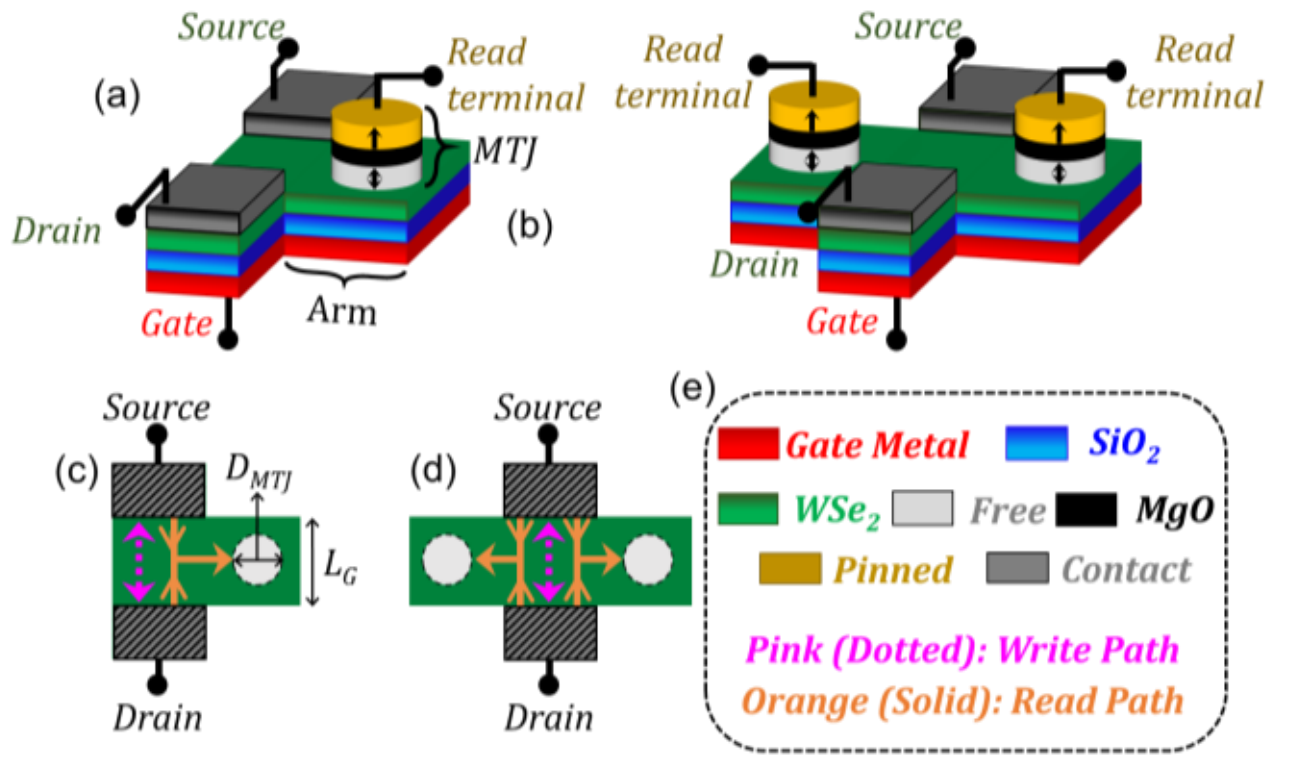

Fig. 6 Proposed (a) single ended (VSH-MRAM) and (b) differential (DVSH-MRAM) non-volatile memory devices. Read and write paths for (c) VSH-MRAM and (d) DVSH-MRAM. (e) Legend.

## V. VSH Effect Based Non-Volatile Memory Design

We propose single-ended and differential variants of non-volatile memories using the VSH effect, namely VSH-MRAM (single-ended) and DVSH-MRAM (differential). We discuss the memory device structures and their characteristics followed by array design in this section.

### A. *Structure and Operation of VSH Memory Devices*

Fig. 6(a, b) illustrates the proposed single ended and differential memory device structures. Single-ended VSH-MRAM consists of only one arm along which the transverse spin current flows. On the other hand, the differential DVSH-MRAM contains two arms for complementary spin current flow. In the single ended design, a PMA MTJ is integrated on top of the arm of the monolayer TMD spin generator as shown in the Fig. 6(a, b), whose free layer (FL) is used for non-volatile magnetic storage. In the differential design, two PMA MTJs storing true and complementary values are integrated on the two arms of the spin generator. The read terminals of the memory devices (connected to the pinned layer of the read MTJs - Fig. 6(a, b)) are used to sense the bit-information stored. By virtue of VSH-based write and MTJ-based read (discussed in detail later), the proposed memory devices feature decoupled read-write paths.

The VSH effect in monolayer $WSe_2$ generates out-of-plane spin current ($I_S$), which interacts with the MTJ through spin torque to switch the FL magnetization (Fig. 2). Since VSH effect leads to the flow of opposite spin currents in divergent directions, the proposed DVSH-MRAM is able to seamlessly store and switch both true and complementary bits. The direction of the charge current ($I_C$) (controlled by the polarity of drain-to-source voltage ($V_{DS}$)) determines the polarization of the spin current ($I_{S+}/I_{S-}$) flowing towards the MTJ(s) (see- Fig. 2 and Fig. 6 (c, d)). When the current flows from the drain to source terminals, $I_{S+}$ flows towards the MTJ in VSH-MRAM and $I_{S+}/I_{S-}$ flow towards the right/left MTJ ($MTJ_{R/L}$) in DVSH-MRAM. These spin currents generate spin torque leading to parallel (P) state in the MTJ of VSH-MRAM, and P and anti-parallel (AP) states in $MTJ_R$ and $MTJ_L$ respectively, in DVSH-MRAM. Current is passed in the opposite direction to store the opposite states. This corresponds to the write operation of the proposed memory devices. For reading the bit-information, we use the resistance difference between the P and AP states of MTJs [9-11]. The read current flows from the source and drain terminals of the transistor to the read terminal of MTJs, in a 'T'/ 'H' shape as illustrated in Fig. 6 (c, d), for VSH/ DVSH-MRAMs. The biasing conditions to achieve this is explained in Section V. Based on the current sensed at the read terminals (which depend on the state of MTJ, P/AP), the bit-information stored is retrieved. It is important to note that VSH-MRAM achieves single-ended sensing using a reference current source, while DVSH-MRAM enables differential sensing leading to higher sense margins and self-referenced operation. These aspects are discussed in detail later.

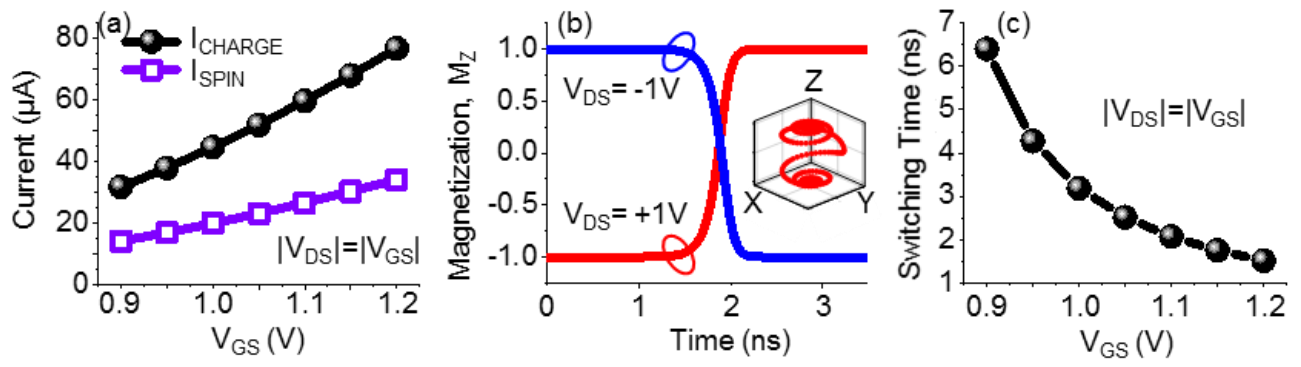

Fig. 7 (a) $V_{GS}$ modulation of generated spin currents (b) Magnetization ($M_Z$) switching for different $V_{DS}$ polarity (inset: magnetization trajectory) (c) Magnetization switching time vs $V_{GS}$.

A unique feature of our devices is the integrated back gate, which enables modulation of the $I_C$, $I_S$ and hence the switching characteristics of the PMA magnets (gate controllability quantified later). While this feature can be appealing for several applications, in this work, we utilize it for compact memory design, as discussed in the subsequent sections.

## B. *Memory Device Characteristics*

As mentioned previously, charge current flowing through the monolayer $WSe_2$ generates transverse spin currents. Fig. 7(a) illustrates the simulated gate voltage ($V_{GS}$) modulated charge and spin current flow. The polarity of $V_{DS}$ determines the polarization for the spin current flowing towards the MTJ resulting in corresponding magnetization switching as illustrated in Fig. 7(b). Since, the gate voltage controls the carrier density in the $WSe_2$ layer, the magnetization switching time is gate controllable as shown in Fig. 7(c). Higher $|V_{GS}|$ corresponds to larger $I_C$ (or $I_S$) which in turn results in smaller magnetization switching time. For our proposed (D)VSH-MRAM devices, we achieve switching time ranging from 3.2ns to 1.5ns for $V_{GS}$=-1.0V to -1.2V. Note, the magnetization switching time for VSH and DVSH-MRAMs remain similar because of the inherent and concurrent generation of the complementary spin currents ($I_{S+}$ and $I_{S-}$) due to VSH effect.

It is important to mention that when $V_{GS}$=0V, the device is OFF and the magnetization state is retained due to the non-volatility of the ferromagnet. To read or change the magnetization state stored, the device has to be turned ON (negative $V_{GS}$). As we discuss later, during read, even though the device is ON, we ensure that no charge current flows from the drain to source terminal (to avoid generation of spin current due to VSH effect), thereby safeguarding the magnetization state from any VSH-induced disturbance. Let us now discuss the proposed memory array design and operation.

## C. *Memory Array Design and Operation*

Utilizing the integrated back-gate of our devices, we propose VSH and DVSH-MRAM arrays which feature access-transistor-less cells by virtue of the integrated gate of the proposed devices (Fig. 8(a, b)). The integrated gates of all the memory cells in the same row are connected to the word-line (WL). The source, drain and the read-ports of all the cells in the same column are connected to bit-line (BL), bit-line-bar (BLB) and sense-line/sense-line-bar (SL/SLB) respectively. The integrated back gate provides selective word access in the array as discussed later. This feature leads to compact layouts as shown in Fig. 8(c, d). The memory operations are discussed next (bias conditions in Table. 1).

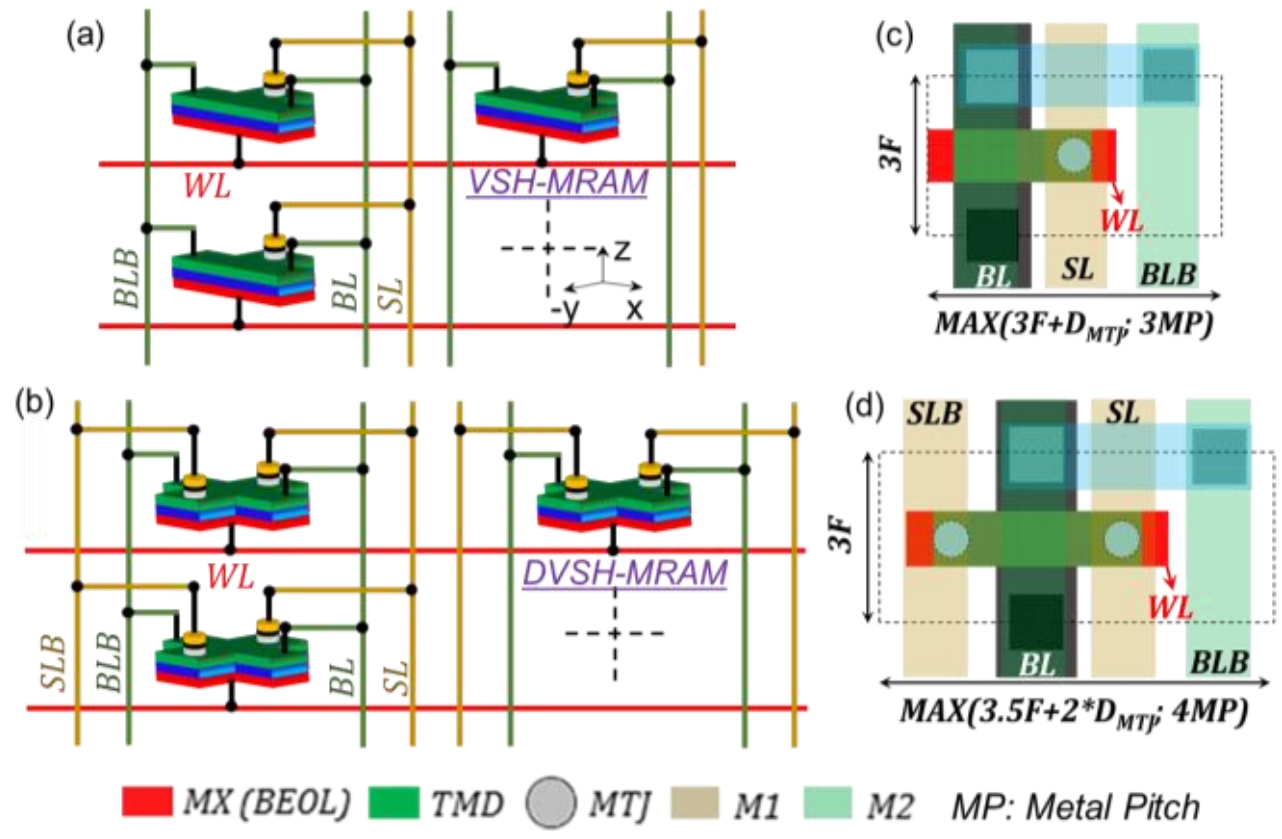

Fig. 8 Memory array architecture and bit-cell layout (with $L_G$=45 nm) for the proposed (a, c) VSH-MRAM and (b, d) DVSH-MRAM.

| Red: Pre-charged | WL | BL | BLB | SL | SLB |
|---|---|---|---|---|---|
| WRITE | 0 | $V_{DD}$/0 | 0/$V_{DD}$ | $V_{DD}$ | $V_{DD}$ |
| READ | 0 | $V_{DD}$ | $V_{DD}$ | $V_{DD}$-$V_{READ}$ | $V_{DD}$-$V_{READ}$ |
| HOLD | $V_{DD}$ | $V_{DD}$ | $V_{DD}$ | $V_{DD}$ | $V_{DD}$ |

*(i) Write:* For writing into the proposed memory cell, we apply 0V to WL of the accessed word (Recall that the proposed devices are p-type). We then assert BLs and BLBs according to the bit-information which is to be stored (note from Section V.A, direction of charge current determines the bit stored). SLs and SLBs are kept pre-charged (and *floating*) at $V_{DD}$ (1.0V). This creates a high impendence path for the charge current to flow through MTJ, avoiding accidental magnetization switching due to STT effect. Now, let us first consider the case where we write bit-'0'. 0V/$V_{DD}$ is applied to BL/BLB in both VSH- and DVSH-MRAMs ($V_{DD}$=1.1V). VSH effect as discussed in Section V.A, flips the FL of MTJ in VSH-MRAM to positive magnetization state ($M_Z$=+1) and the MTJ comes to the P configuration. While for DVSH-MRAM, FL of $MTJ_R$ and $MTJ_L$ flip to positive and negative magnetization states ($M_Z$ =+1 and –1) which brings them to P and AP configurations respectively, corresponding to bit-'0'. On the other hand, for writing bit-'1', $V_{DD}$/0V is applied to BL and BLB, and the VSH effect leads to storage of $M_Z$=-1 in FL of MTJ (AP) of VSH-MRAM and $M_Z$= -1/+1 in FL of $MTJ_R$(AP)/$MTJ_L$(P) of DVSH-MRAM. Note, in DVSH-MRAM, the true bit value is stored in $MTJ_R$ while the complementary bit is stored in $MTJ_L$. After write, all lines are pre-charged to $V_{DD}$. Note, the BLs/BLBs, SLs/SLBs of the unaccessed cells are precharged to $V_{DD}$, while the WLs are driven to $V_{DD}$ to avoid any unintentional $M_Z$ switching. This corresponds to $V_{GS}$=$V_{DS}$=0V in the unaccessed memory devices resulting in insignificant charge/spin current flow (no write disturbance).

*(ii) Read:* For reading the bit-information, we apply 0V to WL and $V_{DD}$ to BLs and BLBs of the accessed word. The SLs and SLBs are driven to $V_{DD}$-$V_{READ}$. This brings the memory devices of the accessed word to the ON state and there exists a read current flow between the sense line(s) and source/drain terminals of the memory cell (due to the voltage difference, $V_{READ}$=0.4V). The read current ($I_{SL}$/$I_{SLB}$) depends on the resistance of the MTJ storing P or AP configuration (as discussed in Section V.A). For VSH-MRAM, $I_P$ is the current sensed at SL when the memory cell stores bit-'0' (parallel configuration of MTJ) and $I_{AP}$ is the current sensed when bit-'1' is stored (anti-parallel MTJ), where $I_P > I_{AP}$. For DVSH-MRAM, $I_P$ ($I_{AP}$) and $I_{AP}$ ($I_P$) are the currents sensed at SL and SLB when the bit stored is '0' ('1'). VSH-MRAMs employs single-ended sensing, where a reference cell current, $I_{REF}$= $(I_P+I_{AP})/2$ is used to compare the current flowing through SL ($I_{SL}$). On the other hand DVSH-MRAM is self-referenced. After the read operation, all lines are pre-charged to $V_{DD}$. Note, similar to the write operation, the BLs/BLBs and SLs/SLBs of the unaccessed cells are precharged to $V_{DD}$ and the WLs are driven to $V_{DD}$ to avoid any disturbances.

*(iii) Hold/Sleep:* During the hold operation, all the lines of the memory array are precharged to $V_{DD}$. This process also ensures minimal energy consumption during charging/dis-

| Table.2 MTJ parameters for (D)VSH and (D)GSH-MRAMs | | |
|---|---|---|
| Parameters | PMA | IMA |
| Free Layer Thickness: $T_{FM}$ | 1.25nm | 1.75nm |
| Saturation Magnetization: $M_S$ *[19]* | 1257.3 emu/cc | 1257.3 emu/cc |
| In-plane/Perpendicular Anisotropy Energy Density: K *[19, 23]* | $K_{\perp}$ = 2.5x10$^6$ erg/cc | $K_{\parallel}$ = 0.84x10$^6$ erg/cc |
| Anisotropy Field: $H_K$ | 3.9k Oes | 1.33k Oes |
| Aspect Ratio | 1 | 2 |
| Volume | Π*15*15*1.25 nm$^3$ | Π*30*15*1.75 nm$^3$ |
| $\Delta_{EB}$ | >50$K_BT$ | >50$K_BT$ |

charging of bit-lines for memory's read/write operations. On the other hand, during the sleep mode, i.e., when the power supply is completely shut down for a long time, all lines (BL/BLB, SL/SLB and WL) are driven to 0V. In both these cases (hold and sleep modes), the non-volatility of the magnetization in FL of MTJ ensures storage of the bit-information even in the absence of any external power supply leading to zero stand-by leakage power.

### *D. Results*

*(i) Array-Level Analysis:* We perform memory array analysis of the proposed VSH/DVSH-MRAMs in comparison with existing GSH/DGSH-MRAMs [14-15]. We consider 1MB array (8 banks, each bank with 1024 rows and 1024 columns) with 32-bit words and evaluate the area, write and read metrics. Iso-energy barrier of ~55$K_BT$ (>10 years of retention [19]) for PMA MTJs in the proposed VSH/DVSH-MRAMs and IMA MTJs in GSH/DGSH-MRAMs is considered for a fair evaluation. This is achieved by tuning the device geometry (MTJ parameters listed in Table. 2). Fig. 9 illustrates the array level comparison of the memory designs.

*(a) Layout (Fig. 8(c, d)):* The proposed VSH/DVSH-MRAMs achieve 66/62% lower bit-cell area compared to GSH/DGSH-MRAMs. This is attributed to the access transistor less array design (see Section V.C; Fig. 8), achieved due to the unique integrated back gate feature. The lower area along with other properties of the VSH effect, enhances the energy efficiencies for memory operations for VSH/DVSH-MRAM as discussed next.

*(b) Write:* The write metrics of the proposed VSH and DVSH-MRAMs remain similar because of the inherent and concurrent generation of $I_{S+}$ and $I_{S-}$ due to the VSH effect (as discussed in Section V.B). However, the same property doesn't hold true for the GSH and DGSH-MRAMs because of different number of access transistors (one and two respectively) driving the write operation (Fig. 1). Our analysis show that VSH/DVSH-MRAMs achieve 59%/ 67% lower write energy (WE) and 50%/ 11% lower write time (WT) compared to the GSH/DGSH-MRAM. This is attributed to two factors. First, the unique generation of out-of-plane spin currents with VSH-effect enables the switching of PMA magnets, unlike GSH effect which can only switch IMA magnets. It is well established that IMA switching is relatively less energy-efficient than PMA switching due to demagnetization fields [19]. Second, lower cell area in the proposed memories results in reduced time and energy consumption for bit-line charging/dis-charging during the write operation.

*(c) Read:* The PMA MTJs in VSH/DVSH-MRAMs exhibit higher resistance due to its smaller area compared to IMA MTJs in GSH/DGSH-MRAMs at iso-energy barrier (Table. 2). Moreover, the $WSe_2$ FET is more resistive than a silicon-based FET used in (D)GSH-MRAMs due to lower mobility. This results in lower sensing currents in VSH/DVSH-MRAMs during the read operation. At the same time, lower area of the proposed memory array due to the integrated back gate feature reduces the bit-line charging/dis-charging energy. Both these factors lead to 74%-77% lower read energy consumption in the proposed memories. However, the lower sensed currents result in 45% lower sense margin for VSH/DVSH-MRAMs compared to GSH/DGSH-MRAMs, at $V_{READ}$=0.4V. At iso-sense margin (achieved by reducing $V_{READ}$ for (D)GSH-MRAM to 0.15V), 35%/ 41% lower read energy is achieved by VSH/DVSH-MRAMs.

With respect to the single-ended VSH-MRAMs, differential DVSH-MRAMs exhibit 50% improved sense margin with a penalty of 64% increase in read energy, attributed to the additional sense-line (SLB) charging energy. However, at iso-sense margin, achieved by reducing $V_{READ}$ of DVSH-MRAM to 0.2V, we observe similar read energies for VSH and DVSH-MRAMs.

*(ii) System-Level Analysis:* With the understanding of thr array-level benefits and trade-offs for the proposed memories, we now evaluate the application-level memory energy benefits of the proposed (D)VSH-MRAMs compared to the existing (D)GSH-MRAMs in the context of (a) general purpose processor and (b) intermittently-powered system.

*(a) General purpose systems:* We evaluate the system-level benefits of the proposed VSH-MRAM and DVSH-MRAM designs when used as an L2 cache (unified memory) in a general-purpose processor. Fig. 10(a) details the system configuration, wherein we design a 2-MB, 8-way set-associative cache using the baseline (GSH-MRAM and DGSH-MRAM) and the proposed (VSH-MRAM and DVSH-MRAM) memory designs. We use gem5 [36], a cycle-accurate architecture simulator, to perform the full-system

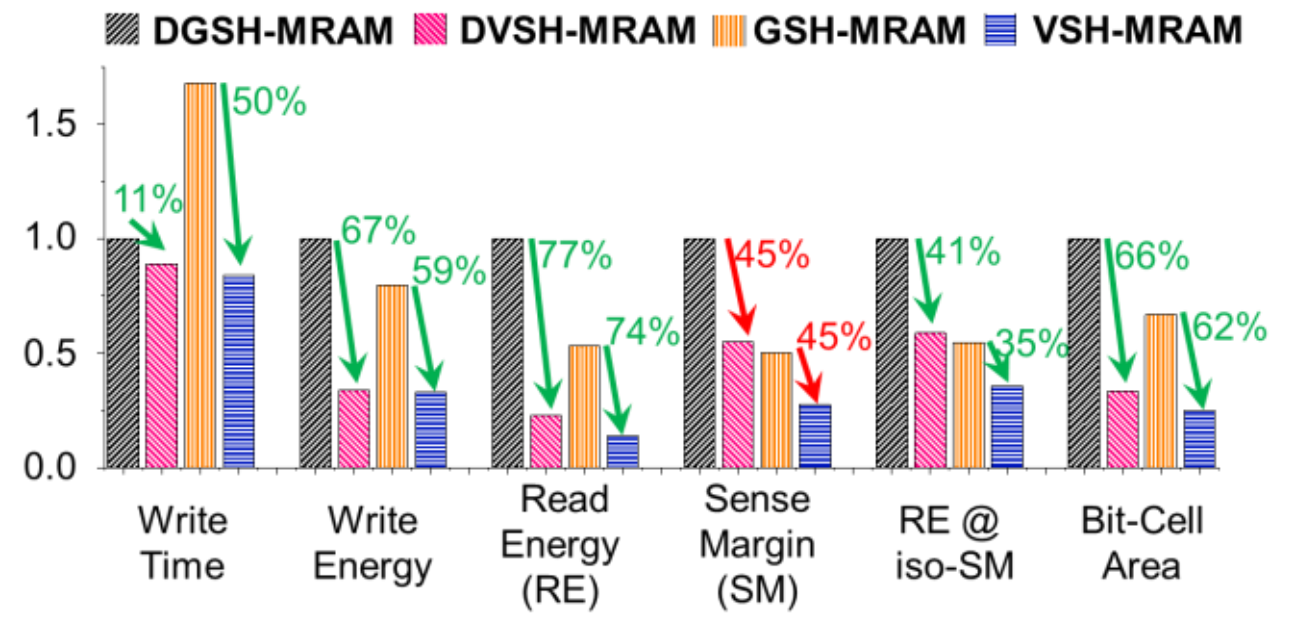

Fig. 9 Array level write-read-layout metric comparison of the proposed VSH/DVSH-MRAMs with GSH/DGSH-MRAMs (normalized). *Note: Iso-SM analysis has been performed individually for single-ended and differential designs.*

| (a) | System Configuration |
|---|---|
| **Components** | **Value** |
| Processor Core | Alpha, Out-of-Order processor, 1core @ 2 GHz |
| L1 I/D cache | 32 KB/64KB, 2 way-set associative, 64B line size |
| L2 unified cache | 2 MB, 8 way-set associative, 64B line size |
| Cache latency | L1 cache: 1 cycle, L2 cache: 8 cycles |
| Memory Design | Iso-sense margin; iso-capacity |

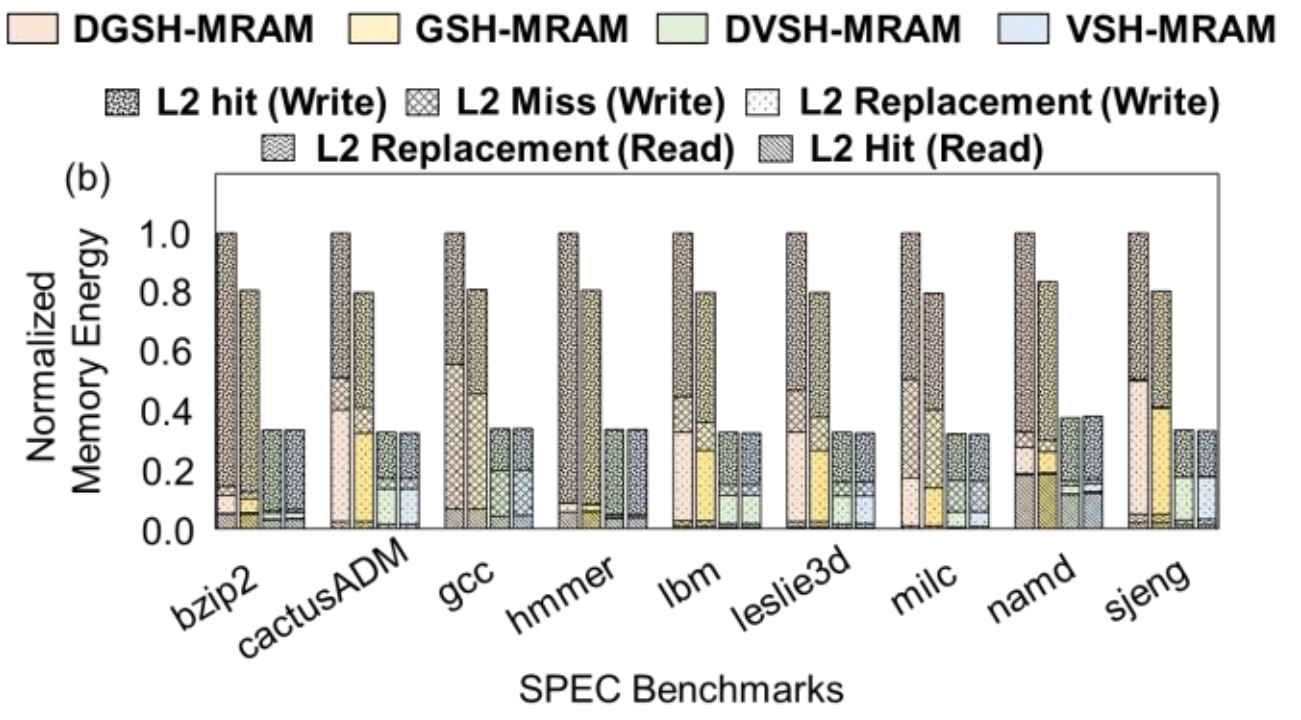

Fig. 10 (a) System configuration used in the general-purpose processor based system analysis (b) Normalized memory energy for various SPEC benchmarks for DGSH, GSH, DVSH and VSH-MRAMs.

simulation and generate memory access traces. We estimate the total L2 cache energy for baseline and proposed designs using the memory traces and the array-level energy results discussed in the previous sub-section.

Fig. 10(b) shows the normalized L2 cache energy for the baseline (GSH/DGSH) and proposed (VSH/DVSH) designs. It also shows the energy consumed by the major L2 cache operations, which are, reads during L2-hits, reads and writes during L2-replacements, and writes during L2-misses and L2-hits. Across a suite of SPEC2K6 benchmarks, VSH-MRAM and DVSH-MRAM exhibit similar L2 cache energy due to similar write and read energies (at iso-sense margin) as discussed in Section V.F. In comparison with DGSH-MRAM and GSH-MRAM, the proposed VSH-MRAMs and DVSH-MRAMs show 2.63-3.14X and 2.19-2.50X reduction in the total L2 cache energy, respectively. Further, the applications (e.g., milc) with a lower read/write ratio show higher benefits. This is because the proposed designs can perform writes far more efficiently compared to the baseline designs.

*(b) Intermittently-powered systems:* Due to the tight energy constraints of intermittently powered systems, we choose the more energy-efficient design for GSH memory for this analysis (single ended GSH-MRAM consumes less energy than the differential design- Fig. 9). Also, for fair comparison, our analysis covers only VSH-MRAM (although both VSH and DVSH MRAMs show similar energy efficiency at iso-sense margin as discussed before). We use a simulation framework shown in Fig. 11(a), similar to the one used in prior works [37]. Our system-level simulations are based on the TI MSP430FR5739 microcontroller-based edge platform running at 24MHz [37] and use a unified 32kB NVM based on the proposed VSH-MRAM (with iso-sense margin of 1.85µA compared to the baseline GSH-MRAM; see Fig. 9) The system is powered using an energy harvesting source that charges a supply capacitor of 10nF. The set of real

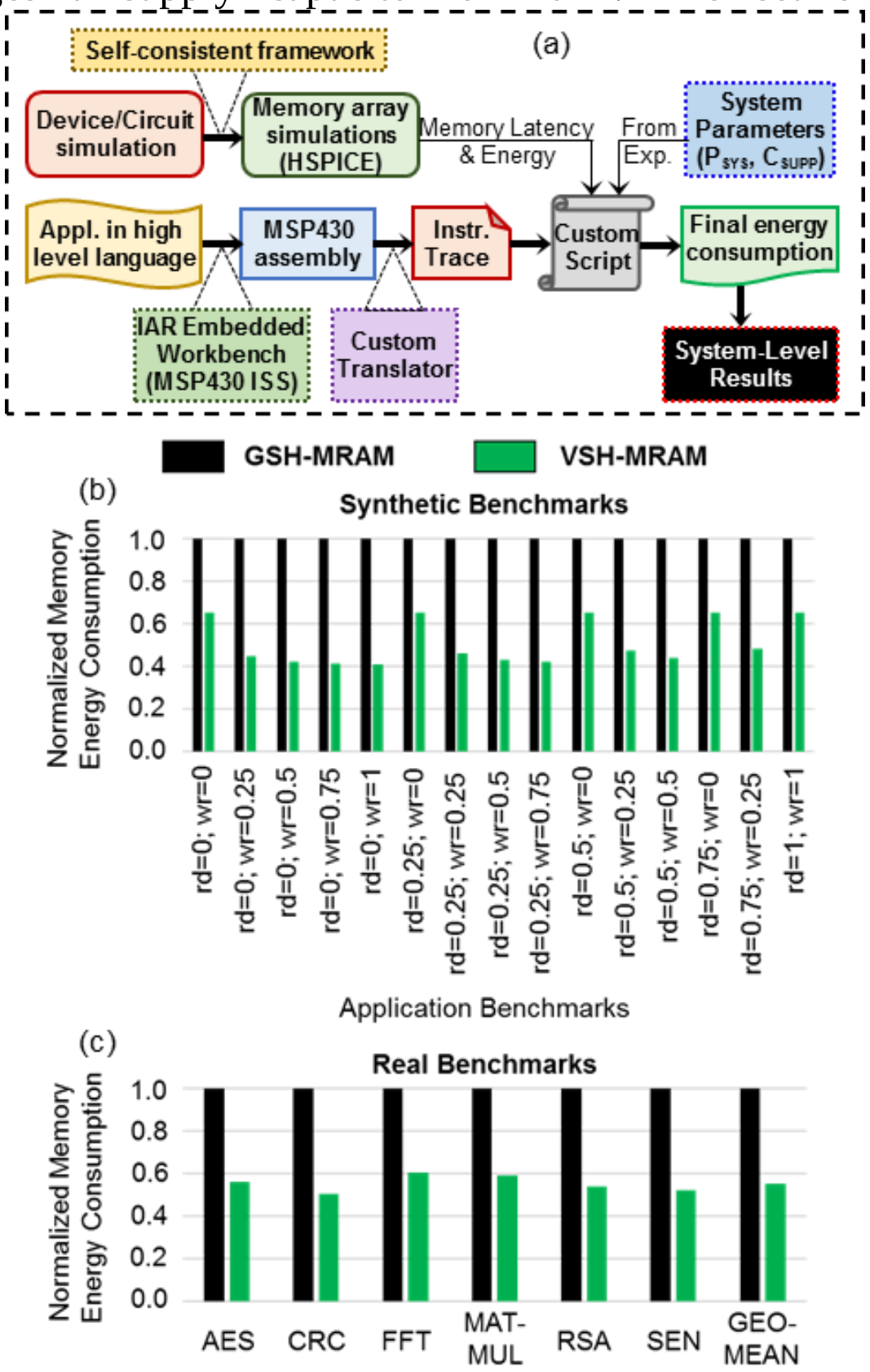

Fig. 11 (a) Simulation framework of IPS to evaluate the memory designs. Normalized system energy consumption of VSH and GSH-MRAM for (b) synthetic and (c) real application benchmarks.

benchmarks are same as that used in [37]. All results discussed below and showcased in Fig. 11(b, c) depict total memory energy consumption for iso-work conditions. Note, the energy numbers in Fig. 11(b, c) are normalized to GSH-MRAM energy consumption.

The energy savings obtained from using VSH-MRAMs compared to GSH-MRAMs depend primarily on the program characteristics, *i.e.*, total number of reads and writes during program execution while executing a specific application. We constructed a set of synthetic benchmarks where we vary the fraction of total memory read and write instructions with a constant checkpoint size of 128B and total number of instructions (100K). Here, the expression {rd:0.25, wr:0.25} represents that 25% of the total instructions are memory reads, 25% are memory writes, and the rest are normal computational operations. In Fig. 11(b), we observe that the proposed VSH-MRAMs achieve significant energy benefits over GSH-MRAMs, ranging from 35% to 59% for a wide spectrum of synthetic memory instructions. This is attributed to the improved read-write energy (at iso-sense margins, see Section V.F). For real application benchmarks, we observe that VHS-MRAMs exhibit energy savings in the range of 40% - 49% (1.66X-1.98X) and 45% (1.80X) on an average compared to GSH-MRAMs (Fig. 11(c)).

## VI. Computation In Memory

In the previous sections, we discussed how the proposed device-circuit design techniques enable single-ended and differential memories which yield significant improvements in area and read/write energies compared to their GSH counterparts. In this section, we go beyond the standard memory operation and utilize the simultaneous true and complementary bit storage of the proposed DVSH-MRAM to enable energy efficient computation in memory (DVSH-MRAM: CiM). As discussed later, by utilizing the multi-wordline assertion (proposed for STT MRAMs in [23]), *natural* and *simultaneous* generation of bit-wise AND and NOR logic functions is achieved in DVSH MRAMs. Utilizing the outputs of these logics in conjunction with other logic gates, we propose a compact compute module to perform computation-in-memory of Boolean logic functions and arithmetic addition (ADD). We also evaluate the proposed single-ended VSH-MRAM for computation in-memory (VSH-MRAM: CiM) based on multi-word-line assertion [23]. To enable computations within the memory array along with standard memory operations, we present a reconfigurable sense amplifier which switches between memory and compute modes as discussed next.

### *A. Reconfigurbale Current Sense Amplifier (RCSA)*

We present a current based reconfigurable sense amplifier which can dynamically switch its operation between differential sensing mode (for memory-read) and single ended sensing mode (for computation in-memory; discussed later). RCSA is designed for the DVSH-MRAM design where the complementary bit-storage can be efficiently harnessed to enable computation in memory. For VSH-MRAM, which is single-ended, we use a standard current-mirror based sense amplifier along with a reference current generation circuit.

The circuit diagram of the RCSA is shown in Fig. 12(a). It contains a pair of core amplifiers (block-A and block-B) based on current-mirroring. During the standard memory-read mode, the two amplifiers are connected together by applying $V_{DIFF}=V_{DD}$ and $V_{DIFFB}=0$. This results in self-referenced differential sensing of the bit stored, resulting in OUT1 and OUT2 that correspond to the currents through SL and SLB.

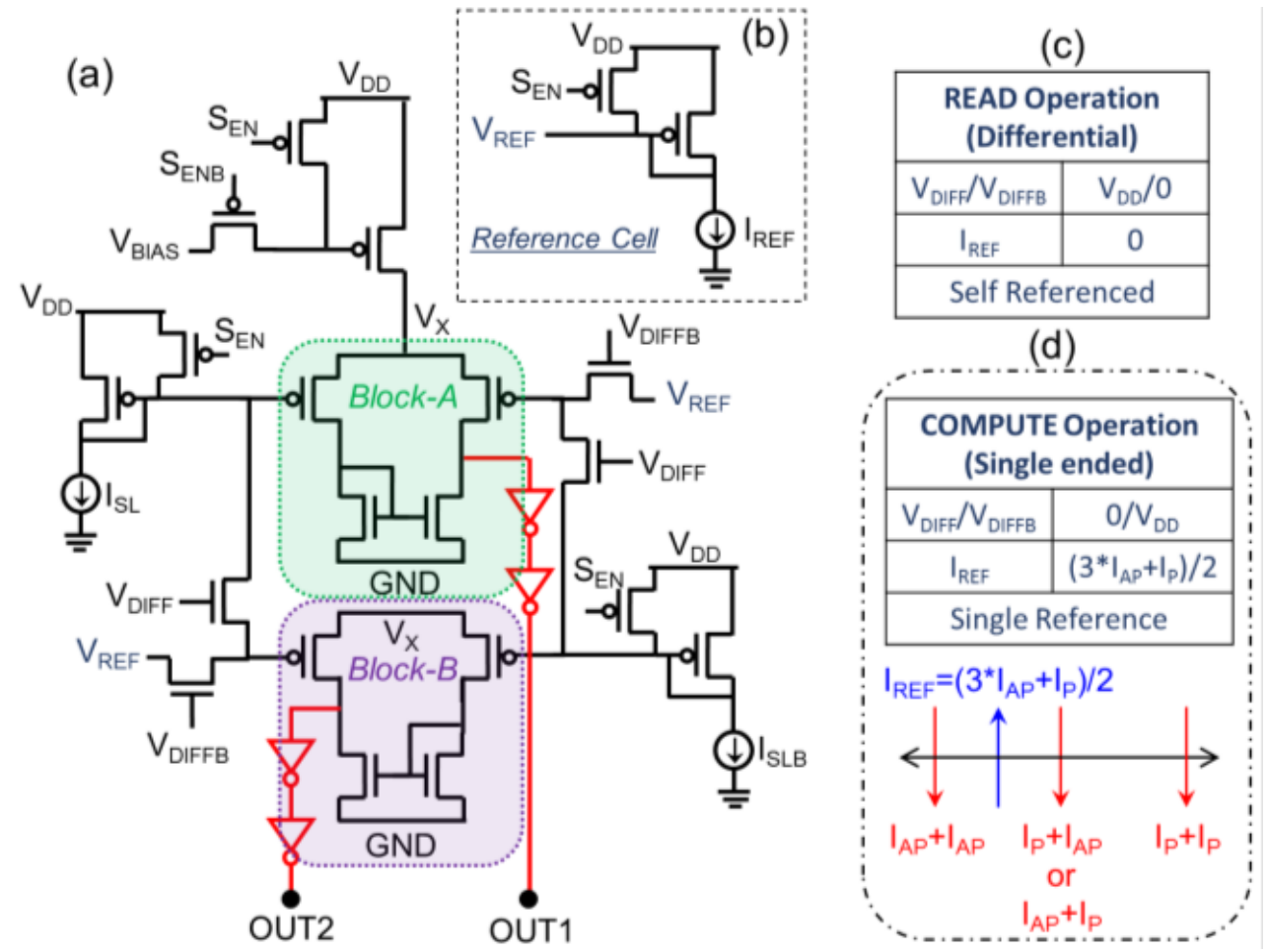

Fig. 12 (a) Reconfigurable current sense amplifier along with (b) the bias conditions for single ended and differential sensing. (c) Reference current used for in-memory compute operations.

The reference generation circuit (Fig. 12(b)) is turned OFF in this mode. On the other hand, during the compute-in-memory mode, where bit-wise computations are carried out individually at SL and SLB (as discussed later), we apply $V_{DIFF\ (B)}$ = 0 ($V_{DD}$) which decouples the two amplifiers. This results in individual single-ended sensing of SL and SLB based on the reference cell current.

## B. Bit-wise AND and NOR logics

The compute operation in the proposed technique is based on the simultaneous assertion of two WLs (Fig. 13(a)) [21-26]. The compute operation follows the same bias conditions as read operation of DVSH-MRAM with RCSA being operated in the single-ended mode ($V_{DIFF}$=0V). The reference current for the single-ended sensing during the compute operations is $I_{REF}=(3^*I_{AP}+I_P)/2$ (Fig. 12(d)). Let us consider two bit-cells storing X: bit-'0' and Y: bit-'1'. When $WL_X$ and $WL_Y$ are asserted (see Fig. 13(a)), the currents through SL, $I_{SL}$ is equal to $I_P+I_{AP}$ (from $MTJ_R$ of X (P) and $MTJ_R$ of Y (AP)) and that through SLB, $I_{SLB}$ is equal to $I_{AP}+I_P$ (from $MTJ_L$ of X (AP) and $MTJ_L$ of Y (P)). Now, since $I_{SL}=I_{SLB}>I_{REF}$, OUT1 and OUT2 are brought to 0V. The truth table for all other input combinations (bit-information stored) is given in Fig. 13(b). Therefore, we *naturally* and *simultaneously* generate bit-wise AND (OUT1) and NOR (OUT2) logic functions at the two ends of the RCSA, with only one reference and without any additional circuitry. This is similar to previous proposals using SRAMs and ferroelectric NVMs for compute-in-memory [21, 22]. The major difference compared to [21, 22] is the use of differential spintronic devices for performing CiM operations and the implementation of the RCSA which enables dynamic switching between *current based* standard memory-read mode and compute-in-memory operation mode. The generated outputs, OUT1 and OUT2 are integrated with the compute module for the computation of other functions as discussed next sub-section.

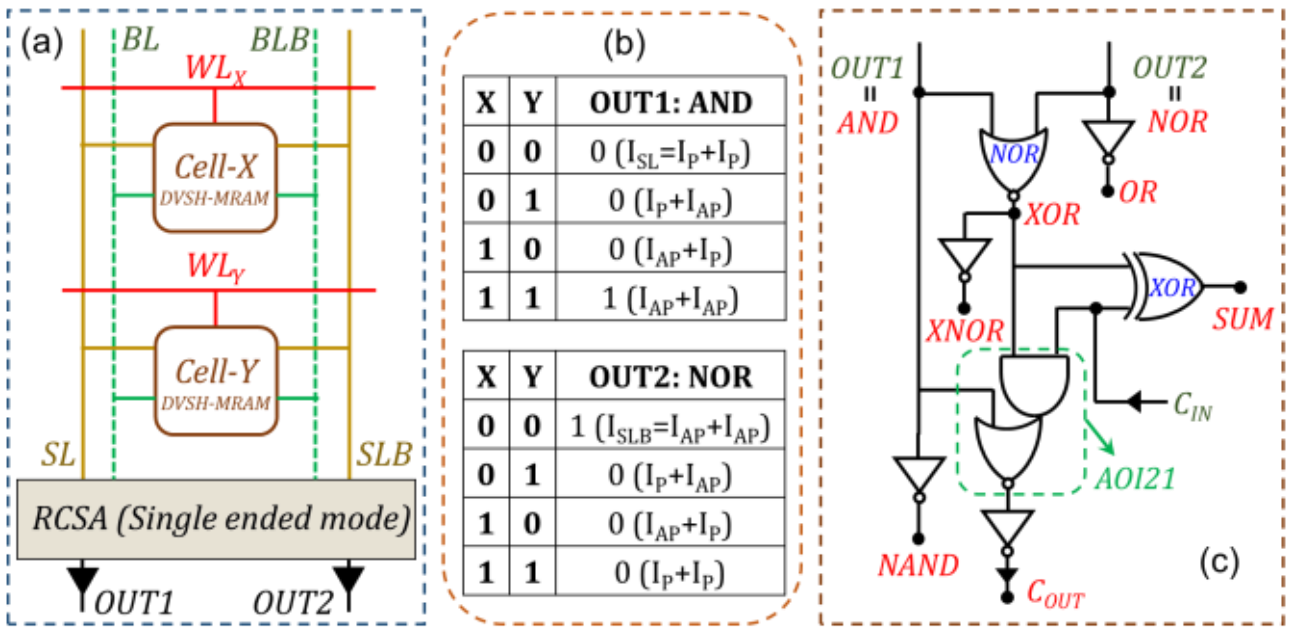

Fig. 13 (a) Example of multi-word line assertion (b) Truth tables for the natively generated AND (OUT1) and NOR (OUT2) functions (c) compute module attached to RCSA for Boolean logic and ADD.

For VSH-MRAM: CiM, we use a single ended current sense amplifier as in [23] and utilize two reference current schemes i.e, $I_{REF}$= $(3^*I_{AP}+I_P)/2$ and $(I_{AP}+3^*I_P)/2$ for achieving the logics AND and OR similar to what has been discussed extensively for STT-MRAMs [23].

The advantages of having a differential memory functionality (like in DVSH-MRAM) compared to single-ended design are (a) the use of only one reference current source for performing the compute operations and (b) natural and simultaneous generation of AND and NOR logic at SL and SLB, as discussed above. These unique attributes lead to significant energy savings which is discussed later.

## C. Compute Module Integrated with the RCSA

In order to realize computing in-memory, which includes bit-wise Boolean operations such as (N)AND, (N)OR, X(N)OR as well as arithmetic operations such as addition (ADD), we present a low power and compact compute module as shown in Fig. 13(c). The naturally generated bit-wise AND and NOR functions of DVSH-MRAM:CiM are simultaneously inverted using standard inverter to compute NAND and OR functions. Using AND and NOR as the input operands for a standard NOR logic gate, we achieve the bit-wise XOR function as shown in Fig. 13(c) which is also simultaneously inverted to achieve the XNOR function. We also implement an in-memory ripple carry adder (RCA) utilizing the bitwise Boolean operations discussed above along with three additional standard logic gates (Fig. 13(c)). The carry-out ($C_{OUT}$) from the previous stage is propagated as carry-in ($C_{IN}$) to the next stage. In our evaluations (next sub-section), we consider a 32-bit word where the $C_{OUT}$ to $C_{IN}$ routing is performed in compute module of the adjacent bit. For VSH-MRAM:CiM, we use the approach proposed in [23].

Next, we evaluate the array and system-level implications of the CiM design for VSH and DVSH-MRAMs. For our baselines, we use the same methodology for compute-in-memory in GSH and DGSH-MRAMs, as discussed in the previous sub-section.

## D. Results

*(i) Array-Level Analysis (Fig. 14):* Similar to the analysis performed in Section V, we evaluate a 1MB array for CiM. During compute operations, due to (a) lower currents through the sense line and (b) lower charging/discharging energy of the bit/sense-lines (similar to read operation discussed in Section V) the proposed VSH/DVSH-MRAM: CiM exhibits is 54%/ 71% lower compute energy consumption for ADD operation when compared to GSH/DGSH-MRAM: CiM design. The lower sensed currents for compute operations are attributed to (a) the higher resistance of the PMA MTJs over IMA MTJs at iso-thermal energy barrier (~55$K_B$T) and (b) higher resistance of 2D TMD channel compared to Silicon FET (see Section V). However, due to this, the sense margin for compute in the proposed VSH/DVSH-MRAMs is 45% lower compared to GSH/D-GSH MRAMs. At iso-sense margin (achieved by reducing $V_{READ}$ (to 0.15V) for GSH/DGSH-MRAM: CiM), the compute energy for ADD operation is 10%/31% lower for the proposed VSH/DVSH-MRAM: CiM. We also compare the proposed differential DVSH-MRAM: CiM design with single-ended VSH-MRAM: CiM and observe that the former achieves 43% lower compute energy consumption at iso-sense margin. This is mainly attributed to (a) natural and simultaneous generation of AND and NOR functions and (b) single current reference for all logic operations.

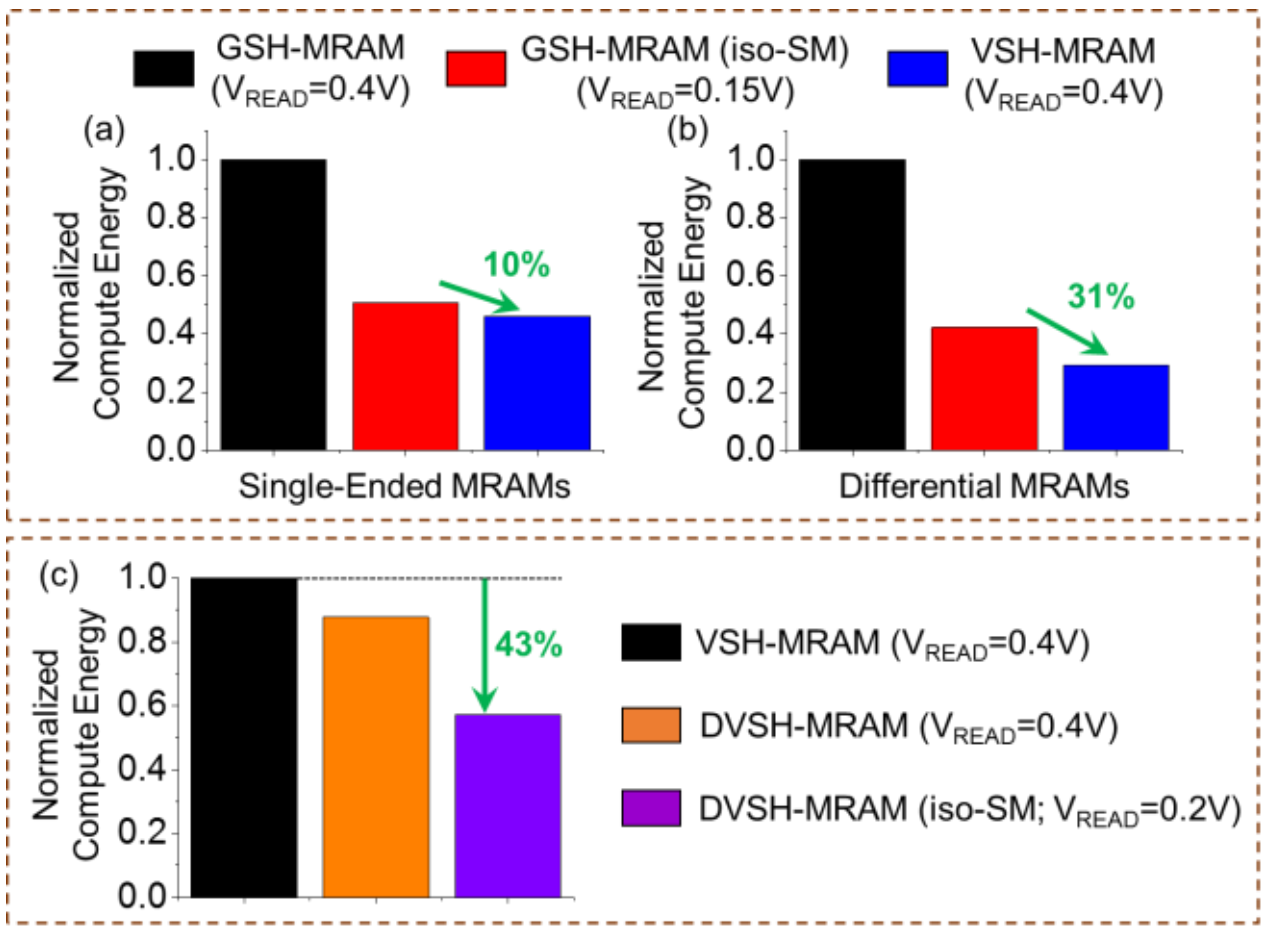

Fig. 14 Array-level normalized compute energy for (a) single-ended and (b) differential memory designs based on VSH and GSH effect. (c) Normalized compute energy consumption of VSH/DVSH-MRAMs.

Due to the superior energy efficiency of differential CiM architectures, we omit the analysis of single-ended VSH/GSH-MRAM CiM design in our system level evaluations (discussed in the next sub-section). Before, we move on, it is important to establish the benefits of CiM over a standard near-memory compute (NMC) architecture where we perform two read operations and then compute in a near-memory logic module. Fig. 15 illustrates the benefits of CiM design over NMC for addition. DVSH-MRAM: CiM exhibits 28% energy benefits compared to DVSH-MRAM: NMC. Compared to DGSH-MRAM: NMC, the proposed DVSH-MRAM:CiM shows 58% energy benefit. Next, we evaluate the proposed CiM design at the system-level.

*(ii) System-Level Analysis (Fig. 16):* We follow the system-level framework used in [23] for our evaluations (see Fig. 16(a)), wherein the proposed DVSH-MRAM: CiM architectures are integrated as a 1-MB scratchpad for the Intel Nios II processor. To expose CiM operations to software, we add custom instructions to the Nios II processor's instruction set, as discussed in detail in [23]. We also extend the Avalon on-chip bus to support CiM operations [23]. Using the array-level results, we estimate the system-level memory energy benefits. We compare DVSH-MRAM:CiM with respect to DGSH-MRAM:CiM, DVSH-MRAM:NMC and DGSH-MRAM:NMC, all with iso-capacity. Further, we design all memories with iso-sense margin of 3.7µA as discussed in Section V (see Fig. 9).

We present the total memory energy benefits for various application benchmarks [23] in Fig. 16 (b). We show all components corresponding to write, read and compute operations for the given application set. We observe that the proposed DVSH-MRAM: CiM achieves total system energy savings of 2.00X to 2.66X over the DGSH-MRAM: NMC, 1.04X to 1.39X over DVSH-MRAM: NMC and 1.45X to 2.57X over DGSH-MRAM: CiM. The benefits primarily arise due to energy- efficient compute operations along with

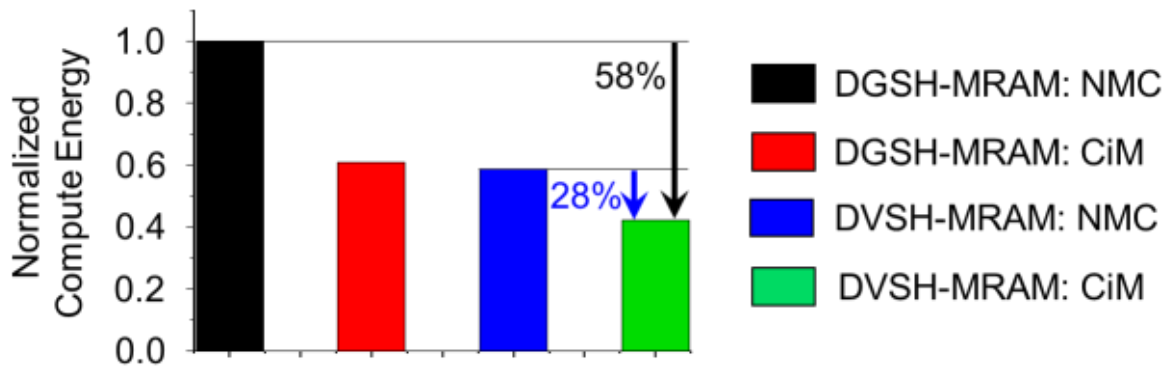

Fig. 15 Array-level normalized compute energy using near-memory compute (NMC) and compute in-memory (CiM) architectures for DGSH-MRAM and DVSH-MRAM.

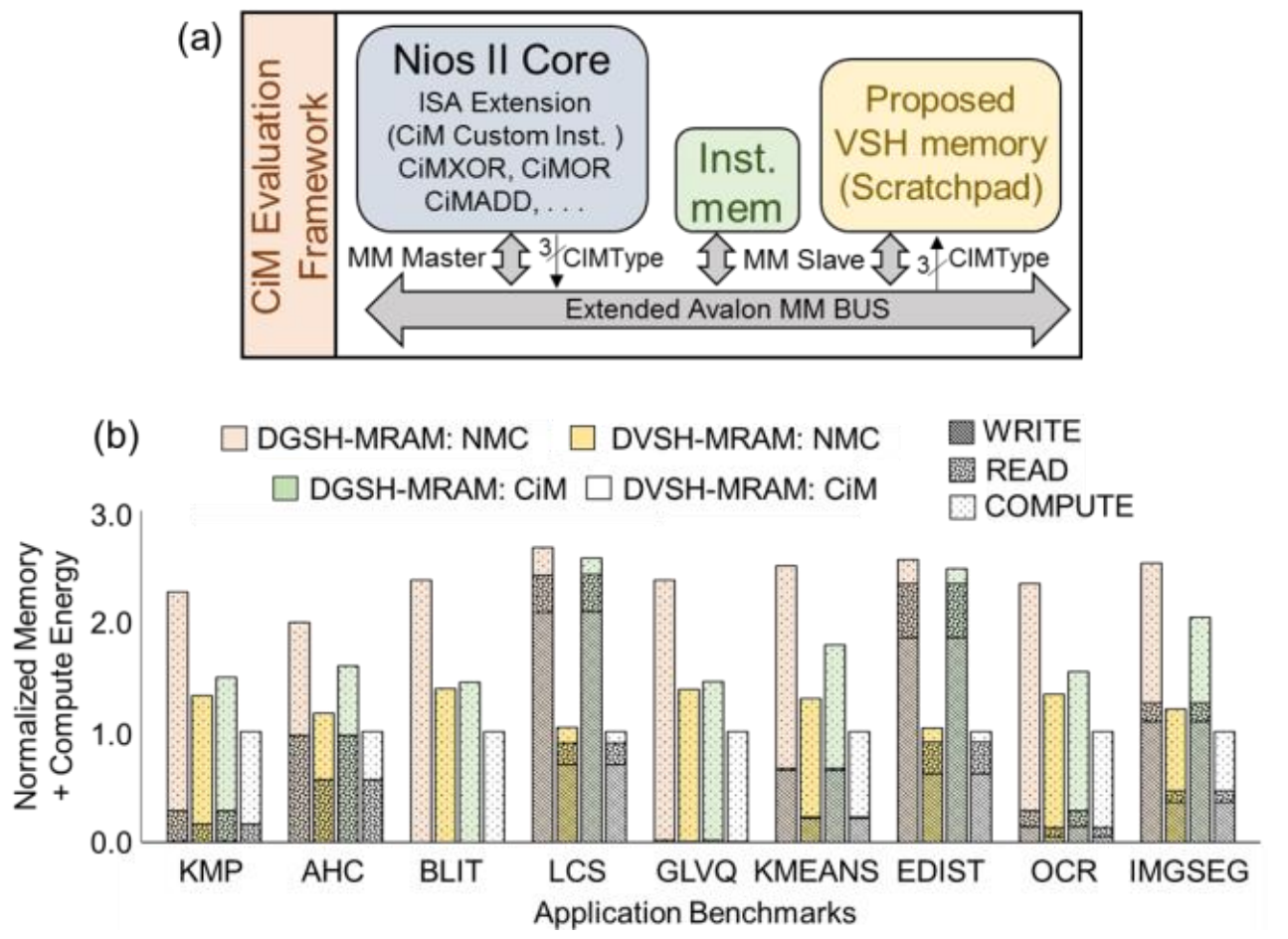

Fig. 16 (a) Simulation framework for system-level evaluation (b) total system energy consumption of the proposed DVSH-MRAM: CiM in comparison with DGSH-MRAM: CiM and DVSH/DGSH-MRAM: NMC for various application benchmarks.

superior read and write operations due to the unique attributes of VSH effect in the propose DVSH-MRAMs. CiM operations reduce memory accesses, bus transfers and processor instructions leading to significant energy savings when compared to the NMC baselines.

## VII. Conclusion

We propose Valley-Spin Hall (VSH) effect based single-ended and differential non-volatile memories with support for low power computation in-memory. The out-of-plane spin generation along with integrated back gate in the proposed memory devices lead to compact array design and high energy efficiency compared to previously proposed Giant Spin Hall (GSH) effect-based memory designs. Both array and system-level analysis was performed for various benchmarks showing the benefits of the proposed VSH effect based memories in terms of low power consumption. We also present a reconfigurable current sense amplifier which can switch its mode of operation between differential sensing for memory read and single-ended sensing for compute-in-memory (CiM). We observe that the proposed differential DVSH-MRAM:CiM design offers improved energy efficiency compared to its single-ended counterpart, at iso-sense margin. We also evaluate our CiM designs at the system-level with respect to GSH effect-based CiM and NMC design as baseline for various application benchmarks and show significantly improved energy efficiency.

## References

[1] J. S. Meena et al., "Overview of Emerging Nonvolatile Memory Technologies" *Nanaoscale Res Lett.*, 9(1), 2014.

[2] Yuan Xie, "Emerging Memory Technologies: Design, Architechure and Applications", *Springer*, 2013.

[3] An Chen, "Emerging Nonvolatile Memory (NVM) Technologies" 45th *Europ. Sol. St. Dev. Res. Conf. (ESSDERC)*, pp.109- 113, 2015.

[4] B. C. Lee et al., "Phase-Change Technology and the Future of Main Memory", *IEEE Micro*, 30(1), 2010.

[5] C. Xu, et al., "Overcoming the challenges of crossbar resistive memory architectures", *International Symposium on High Performance Computer Architecture (HPCA)*, pp. 476-488, 2015.

[6] A. Aziz et al., "Computing with ferroelectric FETs: Devices, models, systems, and applications," in *Proc. Design, Automat. Test Eur. Conf. Exhibit. (DATE)*, 2018, pp. 1289–1298.

[7] S. Hyun, et al., "Dispersion in Ferroelectric Switching Performance of Polycrystalline Hf0.5Zr0.5O2 Thin Films" *ACS applied materials & interfaces* 10 (41), 2018.

[8] S. K. Thirumala and S. K. Gupta, "Gate Leakage in Non-Volatile Ferroelectric Transistors: Device-Circuit Implications" *IEEE Device Research Conference (DRC)*, 2018.

[9] S. P. Park, et al., "Future cache design using STT MRAMs for improved energy efficiency," *Design Auto. Conf.*, pp. 492–497, 2012.

[10] Y. Lee, et al., "Embedded STT-MRAM in 28-nm FDSOI Logic Process for Industrial MCU/IoT Application," *VLSI Symp*., 2018.

[11] O. Golonzka, et al., "MRAM as Embedded Non-Volatile Memory Solution for 22FFL FinFET Technology," *IEDM Tech. Dig*., 2018.

[12] Y. Xie, J. Ma, S. Ganguly and A. W. Ghosh, "From materials to systems: a multiscale analysis of nanomagnetic switching", *Jour. of Comp. Elec.*, 16(4), pp.1201-1226, 2017.

[13] L. Liu, C.-F. Pai, Y. Li, H. W. Tseng, D. C. Ralph, and R. A. Buhrman, "Spin-torque switching with the giant spin Hall effect of tantalum," *Science*, vol. 336, no. 6081, pp. 555–558, May 2012.

[14] Y. Seo, X. Fong, K.-W. Kwon, and K. Roy, "Spin-Hall magnetic random-access memory with dual read/write ports for on-chip caches," *IEEE Magn. Lett*., vol. 6, May 2015, Art. no. 3000204.

[15] Y. Kim S. H. Choday, and K. Roy, "DSH-MRAM: Differential Spin Hall MRAM for On-Chip Memories," *IEEE Electron Device Letters*, vol.34, no.10, pp.1259-1261, Oct. 2013.

[16] C.-F. Pai, L. Liu, Y. Li, et al., "Spin transfer torque devices utilizing the giant spin Hall effect of tungsten," *Appl. Phys. Lett.*, vol. 101, no. 12, pp. 122404-1–122404-4, Sep. 2012.

[17] L. Liu. Et al., "Current-Induced Switching of Perpendicularly Magnetized Magnetic Layers Using Spin Torque from the Spin Hall Effect", *Physical Review Letters*, 109, 096602, 2012.

[18] G. Yu, et al., "Switching of perpendicular magnetization by spin–orbit torques in the absence of external magnetic fields", *Nature Nanotechnology*, 9, pp. 548-554, 2014.

[19] Ikeda, S. et al. "A perpendicular-anisotropy CoFeB-MgO magnetic tunnel junction". *Nat. Mater.* 9, 721–724 (2010).

[20] W. Wulf and S. McKee, "Hitting the Memory Wall: Implications of the Obvious." ACM Comp. Arch. News, 23(1), pp. 20-24, 1995.

[21] S. Jeloka, et al., "A 28 nm Configurable Memory (TCAM/BCAM/SRAM) using Push-Rule 6T Bit Cell Enabling Logic-in-Memory" JSSC, 51(4), pp. 1009–1021, 2016.

[22] S. Thirumala et al., "Non-Volatile Memory utilizing Reconfigurable Ferroelectric Transistors to enable Differential Read and Energy-Efficient In-Memory Computation" *Int. Symposium On Low Power Electronic Design (ISLPED*), 2019.

[23] S. Jain, et al., "Computing in Memory with Spin-Transfer Torque Magnetic Ram" *Trans. On VLSI Systems*, 26(3), pp. 470-483, 2017.

[24] Z. He, Y. Zhang, A. Angizi, B. Gong and D. Fan, "Exploring a SOT-MRAM Based In-Memory Computing for Data Processing", *IEEE Trans. On Multi-Scale Computing Systems*, 4(4), pp. 676-685, 2018.

[25] Z. He S. Angizi D. Fan "Exploring STT-MRAM based in-memory computing paradigm with application of image edge extraction" *Proc. IEEE Int. Conf. Comput. Des.*, pp. 439-446, 2017.

[26] L. Zhang, E. Deng, H. Cai, Y. Wang, L. Torres, A. Sanial, Y. Zhang, "A high-reliability and low-power computing-in-me mory implementation within STT-MRAM" *Microelectronics Journal*, vol. 81, pp. 69-75, 2018.

[27] R. Andrawis, A. Jaiswal and K. Roy, "Design and Comparative Analysis of Spintronic Memories Based on Current and Voltage Driven Switching", *IEEE Trans. on Elec. Dev.*, 65(7), pp. 2682-2693, 2018.

[28] P. Debashis and Z. Chen, "Experimental Demonstration of a Spin Logic Device with Deterministic and Stochastic Mode of Operation", *Scientific Report*, 8, 11405, 2018.

[29] Z. Zhu, et al., "Giant spin-orbit-induced spin splitting in two-dimensional transition-metal dichalcogenide semiconductors", *Phys. Rev. B - Condens. Matter Mater. Phys.* 84, 153402 (2011).

[30] T. Hung, K. Camsari, S. Zhang, P. Upadhyaya and Z. Chen, et al., "Direct Observation of Valley Coupled Topological Current in MoS2" *Science Advances*, 5(4), 2018.

[31] T. Hung, et al., "Experimental observation of coupled valley and spin Hall effect in p-doped $WSe_2$ devices", *arXiv:1908.01396*, 2019.

[32] E. Barre, "Spatial Separation of Carrier Spin by the Valley Hall Effect in Monolayer $WSe_2$ Transistors", *Nano Lett*, 19(2), pp. 770-774, 2019.

[33] S. Suryavanshi and E. Pop, "S2DS: Physics-based compact model for circuit simulation of two-dimensional semiconductor devices including non-idealities", *Jou. of Applied Physics*, 120, 224503, 2016.

[34] K. Y. Camsari, S. Ganguly, and S. Datta, "Modular approach to spintronics," *Sci. Rep.*, vol. 5, Apr. 2015, Art. no. 10571.

[35] C. English et al., "Improved Contacts to MoS2 Transistors by Ultra-High Vacuum Metal Deposition" *Nano Letters*, 16 (6), pp.3824-3830, 2016. DOI: 10.1021/acs.nanolett.6b01309

[36] N. Binkert et al., "The gem5 simulator", *SIGARCH Comput. Archit. News*, 39 (2), pp. 1-7, 2011.

[37] A. Raha et al., Designing energy-efficient intermittently powered systems using spin-hall-effect-based nonvolatile sram". *IEEE Trans. On VLSI Sys. (TVLSI)*, 26(2), pp. 294–307, 2018.